\documentclass[twocolumn, twocolappendix]{aastex631}
\usepackage{amsmath}

\newcommand{\themis}{\textsc{Themis}}

\newcommand{\CC}{C\nolinebreak\hspace{-.05em}\raisebox{.4ex}{\tiny\bf +}\nolinebreak\hspace{-.10em}\raisebox{.4ex}{\tiny\bf +}}
\def\CC{{C\nolinebreak[4]\hspace{-.05em}\raisebox{.4ex}{\tiny\bf ++}}}

\begin{document}
\submitjournal{The Astrophysical Journal}
\title{Autoencoding Labeled Interpolator, \\ Inferring Parameters From Image, And Image From Parameters}

\shorttitle{ALINet: Inferring Parameters from Image \& Image from Parameters}

\author[0009-0003-4620-8448]{Ali SaraerToosi}
\affiliation{Perimeter Institute for Theoretical Physics, 31 Caroline Street North, Waterloo, ON, N2L 2Y5, Canada}
\affiliation{Department of Physics and Astronomy, University of Waterloo, 200 University Avenue West, Waterloo, ON, N2L 3G1, Canada}

\author[0000-0002-3351-760X]{Avery E. Broderick}
\affiliation{Perimeter Institute for Theoretical Physics, 31 Caroline Street North, Waterloo, ON, N2L 2Y5, Canada}
\affiliation{Department of Physics and Astronomy, University of Waterloo, 200 University Avenue West, Waterloo, ON, N2L 3G1, Canada}
\affiliation{Waterloo Centre for Astrophysics, University of Waterloo, Waterloo, ON N2L 3G1 Canada}

\begin{abstract}
    The Event Horizon Telescope (EHT) provides an avenue to study black hole accretion flows on event-horizon scales. Fitting a semi-analytical model to EHT observations requires the construction of synthetic images, which is computationally expensive. This study presents an image generation tool in the form of a generative machine learning model, which extends the capabilities of a variational autoencoder. This tool can rapidly and continuously interpolate between a training set of images and can retrieve the defining parameters of those images. Trained on a set of synthetic black hole images, our tool showcases success in both interpolating black hole images and their associated physical parameters. By reducing the computational cost of generating an image, this tool facilitates parameter estimation and model validation for observations of black hole system.
\end{abstract}
\section{Introduction}\label{sec:intro}
The Event Horizon Telescope (EHT) is a very long baseline interferometer \citep{2019BAAS...51g.256D, Akiyama_2019} created to observe black holes on event-horizon scales, making it possible to study the astrophysical and gravitational processes in the vicinity of black holes \citep{Broderick_2009, Broderick_2011, Broderick_2014, Broderick_2016, Volkel:2019muj}. So far, Sagittarius A* \citep{SagA_paper1} and Messier 87* \citep{1906.11238} have been imaged by EHT, revealing a non-uniform, variable, and complex structure for the emission regions around these black holes. The images provided us with information about the temperature profile and magnetic fields present in the accretion disks \citep{SagA_paper5, M87_paper5}. They have also provided new ways to gain insights about the metric around the black hole \citep{Johannsen_2016, Broderick_2014, m87_paper6, Kocherlakota_2020, SagA_paper6, salehi2023photon, Broderick_Salehi}, black hole's spin \citep{0508386, Broderick_Loeb_2006a, Broderick_2009, Broderick_2011, Broderick_2016}, and mass \citep{1906.11238, SagA_paper4}, and the dynamics of the accretion process \citep{Broderick_Loeb_2005, Broderick_Loeb_2006a, Tiede_2020, Gerogiev_2022, Broderick_2016_dynamics, Broderick_2022, Ni_2022}. 

However, the images are representations of the EHT's primary data products, visibilities. The visibilities are related to the on-sky image via a Fourier transform; in principle, complete information about the visibility map is equivalent to a complete source image \citep[see, e.g.,][]{Thompson2017}. In practice, visibilities are measured for only a sparse set of spatial frequencies, determined by the sky-projected separation between EHT stations, and therefore do not produce a unique image.  Forward modeling approaches avoid this complication by eschewing image comparisons altogether, making the comparison to the EHT observed visibilities directly  \citep[see, e.g.,][]{m87_paper6,SagA_paper4,Broderick_2009}. These methods additionally benefit from the well-defined (very nearly Gaussian) and, important for likelihood construction, independent uncertainties on the visibilities. Choosing the underlying set of models specifies how the gaps in the visibility data are addressed, reducing and potentially eliminating altogether the ambiguity these gaps introduce.  Fitting physically motivated models, produced from models of the orbiting plasma and subsequent gravitational lensing within the Kerr spacetime, allow direct inferences on the physical properties of the black hole itself and the properties of the accretion flow.

Radiatively inefficient accretion flows (RIAFs) form one such class of physical models, appropriate for the low-accretion rates in M87* and Sgr A* \citep{Yuan_Narayan_2014, Broderick_Loeb_2006a, Broderick_2009}.
These accretion models are constructed to explain the apparent low radiative efficiency of very sub-Eddington sources (like the horizon-scale EHT targets) and are characterized by a geometrically thick, nearly virialized disk.  Based on semi-anlaytic models of the stationary state, \citet{BL06} and \citet{Broderick_2011} developed a set of simplified approximate models appropriate for analysis of mm-VLBI imaging \citep{Broderick_2009,Broderick_2011,Broderick_2016}, in which the plasma density, temperature and  magnetic fields are expressed by radial power laws and vertical Gaussian profiles.  A generalized set of RIAF models, wherein the power-law indexes, vertical scale height, and orbital velocity become free parameters were presented in \citet{Broderick_2020}.
Model images are produced by ray tracing, a process in which at each image pixel on a distant screen photon trajectories are traced backward in time and the equations of relativistic radiative transport are solved \citep{Gold_2020, PhysRevD.94.084025,BroderickBlandford2003,BroderickBlandford2004}.

For the purpose of extracting inferences about source parameters, EHT data is fit within Bayesian parameter estimation frameworks \citep{Broderick_2020}. This is done by sampling a model likelihood such that the data provides direct posteriors on the model parameters. In the case of RIAFs, the parameters, such as black hole spin, mass, and inclination angle, describe physical attributes of the system. However, due to the fact that the equations of polarized radiative transfer need to be solved at each sampling step to generate images, this process is computationally expensive. Even in the case of semi-analytical models, e.g., RIAFs, 
that do not require expensive simulations to generate the predicate plasma distributions
\citep{Broderick_Loeb_2006a, Pu_2018},
producing the $10^9$ images typical of a single EHT data analysis is difficult, and repeating this procedure for the $10^2$ times to assess model-bias and systematic error is intractable even with GPU acceleration and efficient parallelization \citep[see, e.g.,][]{Broderick_2020, Tiede_2020}.
However, variations in the RIAF model parameters induce continuous changes in the resulting images, motivating the use of interpolation schemes to reduce the cost of image generation. Because the relationship between model parameter and images is complicated, we require a nonlinear interpolator.

In this paper, we create a nonlinear interpolation tool based on the principles of generative models in machine learning. This tool, named autoencoding labeled interpolator network (ALINet), is a new type of variational autoencoder (VAE) that reduces the dimensionality of an input image and encodes the information in the image in a lower dimensional space, i.e., the latent space \citep{1312.6114}. Afterwards, the decoder interpolates each point in this latent space uniquely and continuously to one image. Sampling this latent space is equivalent to sampling the space of all possible images, but much faster than solving radiative transfer equations each time an image is to be generated.

However, in a traditional VAE the latent space parameters have no easily-interpretable physical meaning. Since each image has its own unique set of physical parameters, there is a one-to-one mapping between an image and its corresponding parameters.\footnote{If this is not true, the parameters recovered from EHT would also not be unique.} Hence, if a set of latent variables can represent an image, they must also through a well-behaved function, represent the physical parameters representing that image. In an ALINet, a second decoder branch provides a continuous map to the physical parameters associated with the image. When the latent space is sampled, the images and associated parameters are generated quickly and the posteriors of the latent variables are convertible to posteriors of the physical quantities. Finally, we trained a second neural network to find an approximate inverse mapping from the physical parameters to the latent parameters. This network is then used to invert the entire process, i.e., we can recover the images by  inputting physical parameters in the second network. Using its output, i.e., the latent parameter distributions, as the input to the ALINet image decoder we can recover the underlying image described by these parameters.

In \autoref{sec:VAE Interpolation} we describe how a traditional VAE works and discuss why it is not suitable for interpolation. In \autoref{sec: ALI} we elaborate on ALINet and explain how it can interpolate images and their corresponding parameters simultaneously and accurately. We also show that the inverse network can interpolate images given the corresponding physical parameters. In \autoref{sec: Experiments} we validate our interpolator by evaluating it on MNIST and testing it against RIAF black hole images. Finally, in \autoref{sec: conclusions} we conclude.

\section{An Introduction to VAE Methods} \label{sec:VAE Interpolation}

Here we review the motivation for and general properties of VAEs.  In doing so, we make use of a number of terms as they are used in the machine learning literature, which while related to similar concepts in astronomy and astrophysics, occasionally have more narrow technical meanings in our context.  To facilitate the translation and avoid confusion, we have collected a summary of such terms in \autoref{table: terminology defs}, where relevant terms, notations, and their definitions may be found.

\begin{deluxetable*}{lccB}
\tabletypesize{\scriptsize}
\tablewidth{0pt} 
%\tablenum{1}
\tablecaption{Terminology definitions \label{table: terminology defs}}
\tablehead{
 \colhead{terminology} & \colhead{notation}& \colhead{definition}
}
\startdata 
physical parameters & $y$ &  values that are the underlying explanations of the images\\
model parameters & $\theta\,
\,\&\,\, \phi$ &  weights \& biases learnable by the machine learning model\\
latent parameters & $z$ & output values resulting from sampling the distributions output from the encoder of a VAE\\
data & $D$ &  inputs to the machine learning model\\
latent prior distribution & $p(z)$ &  a priori distributions assumed for the latent parameter values\\
surrogate distribution/latent distributions & $q_\phi(z|D)$ &  
probability distributions produced by the encoder, ideally should be equal to $p(z|D)$\\
posterior distribution & $p(z|D)$ &  assuming input $D$, probability that sampling the latent distributions gives latent parameters $z$\\
likelihood & $p(D|z)$ &  probability that the data $D$ is the underlying explanation of latent parameters $z$\\
\enddata
\end{deluxetable*}
\subsection{Variational Autoencoder}\label{subsec:VAE}
A comprehensive introduction to VAEs can be found in \citet{1312.6114}; here we only summarize their key features. An autoencoder is comprised of an encoder, which is a parametrized nonlinear map from the image space to a much smaller-dimension latent space, and a decoder, which is a parameterized inverse map, from the latent space back to the image space. The training of an autoencoder is selecting the parameters, or weights, that characterize the encoder and decoder, to maximize the fidelity of these maps. Different choices may be made regarding the choice of integral operations and dimension of the target latent space.

A VAE differs from a standard autoencoder in that it also enforces a notion of continuity within the latent space: nearby positions in the latent space are forced to produce nearby images. The VAE achieves this continuity by encoding an image to distribution functions, as opposed to just a few numbers. We further impose the condition that these distributions should behave a certain way, e.g., we enforce them to look like Gaussian distributions with zero mean and standard deviation of one. Finally, the encoder outputs a set of latent parameters $z$ which then are input to the decoder. The decoder then reconstructs and outputs an image that, ideally, is identical to the image that was input to the encoder.

The goal of the VAE then is three-fold: (a) learn how to encode a high-dimensional image to a low-dimensional representation in its encoder. (b) learn to reconstruct the image from the latent representation of an image. (c) Do (a) and (b) while keeping the latent space distributions as close to the desired latent-variable prior distributions as possible.
The VAE accomplishes this goal by minimizing a loss function which is called the Evidence Lower Bound (ELBO),
\begin{equation}
    \begin{aligned}
        \text{ELBO}(\phi, \,\theta) = &-\underbrace{E_{z \sim q_\phi(z\,|\,D)}[\log\langle p_\theta(D\,|\,z) \rangle]}_\textrm{reconstruction error} \\&+ \underbrace{D_{KL}[q_\phi(z\,|\,D)\,\,||\,\,p(z)]}_\textrm{regularization},
    \label{eq:ELBO}
    \end{aligned}
\end{equation}
where $E$ denotes expectation value and the subscript of the expectation value, i.e., $z \sim q_\phi (z|D)$, denotes the variable for which the expectation value is being calculated \citep{Neal1998AVO}. That is,
\begin{equation}
    {E_{z \sim q_\phi(z\,|\,D)}[\log\langle p_\theta(D\,|\,z) \rangle]} = \int dz q_\phi(z | D) \log \langle p_\theta (D | z)\rangle,
\end{equation}
where $\phi$ and $\theta$ represent the set of parameters (weights and biases) of the encoder and the decoder, respectively. $D_{KL}$ is the Kullback–Leibler divergence, which is a measure of how close a probability distribution is to another \citep{1404.2000}. The term containing the KL-divergence is responsible for keeping the latent distributions compact. $D$ is representative of the available data or images, $z$ denotes the latent space variables or latent parameters, $p(z)$ is the prior on the distribution of latent parameters, $p_\theta(D|z)$ is the likelihood of D occurring assuming the latent values z, $q_\phi(z|D)$ is the ``surrogate distribution'' which the VAE should learn and make as close to the posterior on the latent variables $p(z|D)$ as possible while keeping the latent space distribution compact. Note that learning the surrogate distribution and keeping the distribution compact are done in the encoder, while the decoder learns the likelihood $p(D|z)$ which leads to the decoder learning to reconstruct the image from the latent space variables. The surrogate distribution is kept compact by enforcing it to be as close to the prior distribution $p(z)$ as possible. The prior distribution is usually assumed to be a Gaussian of the form $p(z) = \mathcal{N}(0, 1)$ which denotes a Gaussian distribution with zero mean and variance one.

If the image dimension is significantly larger than the latent space dimensions, reconstruction loss in \autoref{eq:ELBO} dominates and the regularization loss is insignificant in comparison. This large difference in the latent dimension and the image dimension therefore would impair the ability of the encoder to keep the latent distributions compact and close to the priors. This problem can be resolved if a $\beta$-VAE is used instead \citep{Higgins2016betaVAELB}, where the difference is that the loss function takes the following form:
\begin{equation}
    \begin{aligned}
        ELBO(\phi, \,\theta) = &-\underbrace{E_{z \sim q_\phi(z\,|\,D)}[\log\langle p_\theta(D\,|\,z) \rangle]}_\textrm{reconstruction error} \\&+ \beta \times \underbrace{D_{KL}[q_\phi(z\,|\,D)\,\,||\,\,p(z)],}_\textrm{regularization}
    \label{eq:ELBO beta-VAE}
    \end{aligned}
\end{equation}
where $\beta$ is a hyperparameter that scales the KL-divergence term of the loss function, i.e., how much emphasis is put on keeping the surrogate distribution close to the prior. The value of $\beta$ is chosen such that the regularization loss is comparable to the reconstruction loss.

\subsection{VAE Interpretation}\label{subsec:VAE Interpret}
A problem with VAEs is that the distributions learned by the autoencoder in the latent space are not readily interpretable. This problem becomes more challenging to deal with as the complexity of a neural network is increased. \citep{Towell,7299155,14541,1703.00810}. What is learned by the machine is not necessarily the physical parameters we want it to learn. However, since the difference in the images is caused by the physical parameters, there should exist a mapping from the latent parameters to the physical parameters. 

Three approaches have been used previously, the first and simplest  is to manually generate the mapping. One of these manual efforts is to plot the latent parameters vs. physical parameters and fit a function to find the mapping between them \citep{PhysRevE.96.022140}. However, as the number of variables in the latent space increases, any interpretation of the actual meaning of each variable becomes more difficult. 

The second approach is a more automatic effort of interpretation. One of these efforts was made by \citet{PhysRevB.96.184410} to reconstruct the decision function of a neural network as a function of the input. However, this method is problem-dependent, it requires one to know the underlying behavior of the target function beforehand and needs human intuition and fine-tuning at each step of the way to yield a meaningful result. 

A more popular method is to take advantage of a network architecture called conditional VAE (C-VAE), where the VAE is conditioned on the physical parameters so that the decoder would accept the physical parameters as input \citep{NIPS2015_8d55a249}. The problem with this method is that even though the decoder would be conditioned on the physical parameters, much like the original VAE, there still exists a latent space that is not readily understood and over which we have no control, i.e., we cannot readily manipulate this latent parameters to make them have a meaning that is understandable by humans. The extra degrees of freedom in a C-VAE have the potential to make the model less accurate. To avoid the extra degrees of freedom, we make use of the universal approximation theorem \citep{Cybenko1989ApproximationBS, lu2017expressive} and relegate the task of finding the mapping to the VAE itself.

\section{Autoencoding Labeled Interpolator Network (ALINet)}\label{sec: ALI}
The architecture we implemented can be found in \autoref{fig: 200k architecture}.
In this new structure, which we name autoencoding labeled interpolator network (ALINet), we stitch the labels (physical parameters) to the VAE using a new neural network branch connected to the latent parameters on the decoder side, whose sole job is to find the mapping from the latent parameters to the physical labels. The loss function has, therefore, three components this time, 
\autoref{eq:ELBO-ALI}.
\begin{equation}
    \begin{aligned}
        ELBO(\phi, \theta) = &-\underbrace{E_{z \sim q_\phi(z|D)}[\log\langle p_\theta(D|z) \rangle]}_\textrm{reconstruction error} \\&+ \beta \times \underbrace{D_{KL}[q_\phi(z|D)||p(z)]}_\textrm{regularization}
        \\&+ \alpha \times \underbrace{\sum_i(y_i - \hat{y}_i)^2.}_\textrm{
    SSE}
    \label{eq:ELBO-ALI}
    \end{aligned}
\end{equation}

Note that in our model, $\phi$ still represents the learnable parameters in the encoder, but $\theta$ represents all the learnable parameters in both branches of ALINet decoder, $y_i$ and $\hat{y}_i$ are the truth and network output for the physical parameters, and $\alpha$ is a hyperparameter that tunes the strength of the extra term in the loss function, i.e., how much emphasis is placed on predicting the physical parameters. This hyperparameter is required to make the sum squared error (SSE) loss comparable to the reconstruction loss.\footnote{Since the dimensions of the images are significantly bigger than the number of parameters, reconstruction loss tends to be considerably larger than the parameter SSE loss.} Depending on the accuracy of the parameter that is required, $\alpha$ may be chosen to make SSE losses smaller, comparable, or larger than those associated with the image recovery.

\subsection{ALINet As A Parametric Interpolator}
The ALINet decoder is a mapping from latent parameters to an image described by those latent parameters (output of the first branch), and its corresponding physical parameters (output of the second branch). 
As it is, this already provides a parametric interpolator on the physical parameter space: new images and accompanying physical parameter sets may be generated by sampling the latent space. This is possible in ALINet by virtue of the continuity of the VAE and the addition of the second decoder branch.

Creating an image directly from given input physical parameters can be facilitated by taking advantage of the inverse network discussed in the following section.

\subsection{Inverse Network}
The ALINet architecture in \autoref{sec: ALI} can also retrieve physical parameters given an image. To generate an image given physical parameters, we train an inverse neural network (InvNet) that learns the mapping from physical parameters to the latent distributions. These latent distributions then will be fed to the ALINet decoder to create images. The loss function for this network, therefore, is,
\begin{equation}
    \text{Loss} = \alpha \times \sum_{i} (\hat{\mu}_{\phi, i} - \mu_{\phi, i})^{2} + \sum_{i} (\log(\hat{\sigma}_{\phi, i}^2) - \log(\sigma_{\phi, i}^{2}))^2, 
    \label{Eq: InvNet}
\end{equation}
where $\mu_\phi$ and $\sigma_\phi$ are the means and standard deviations of the latent distributions representing the physical parameters that were learned by the ALINet encoder, variables with hats represent the predictions of the network, and $\alpha$ is a multiplier to adjust the emphasis of the loss function.

The InvNet+ALINet, therefore, can be used to generate images from input parameters. First, the input parameters are given to InvNet. The output of InvNet is then a set of latent parameter distributions. These latent distributions are then fed to the ALINet decoder, where the decoder recovers the image associated with the input latent distributions. Hence, by specifying the input parameters in InvNet+ALINet, we can find the image associated with those parameters.

\section{Numerical Experiments}\label{sec: Experiments}
Here we present applications and their validation of ALINet and associated architectures. 

\subsection{MNIST}\label{subsec: MNIST expr}
\begin{figure*}[ht]
\centering
\includegraphics[width=1\textwidth]{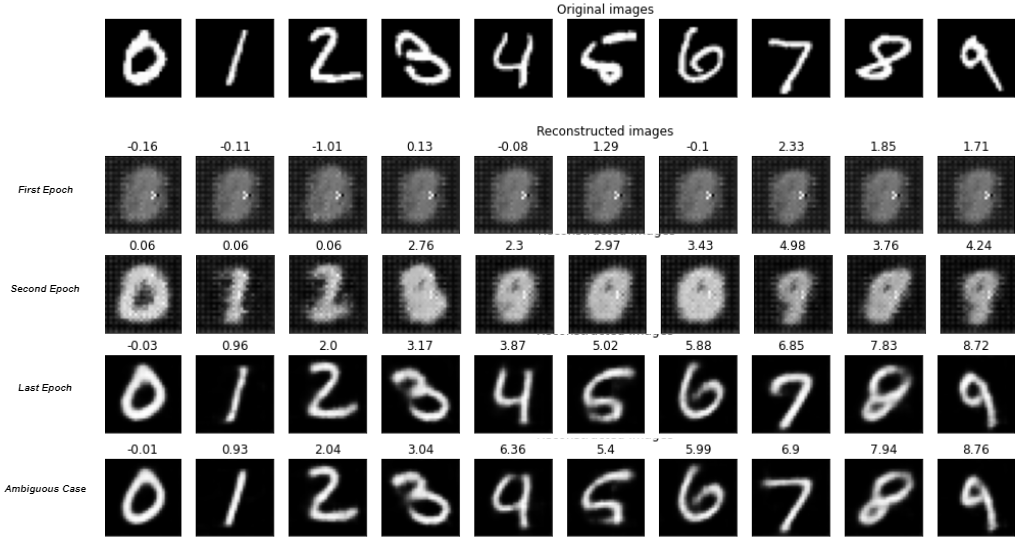}
\caption{
Recovered images and parameters from ALInet for 10 randomly chosen test images after training on the MNIST data set. One test is shown for each digit after one epoch (2nd row), two epochs (3rd row), and 50 epochs (4th row) of training.  For comparison, test image truths are shown at the top.  The bottom row shows an ambiguous case (5th column from the left, in which a 4 and 9 are confused) in both the recovered image and digit value ($(4+9)/2=6.5$).
\label{fig: MNIST epochs}}
\end{figure*}

\begin{figure}
    \gridline{\fig{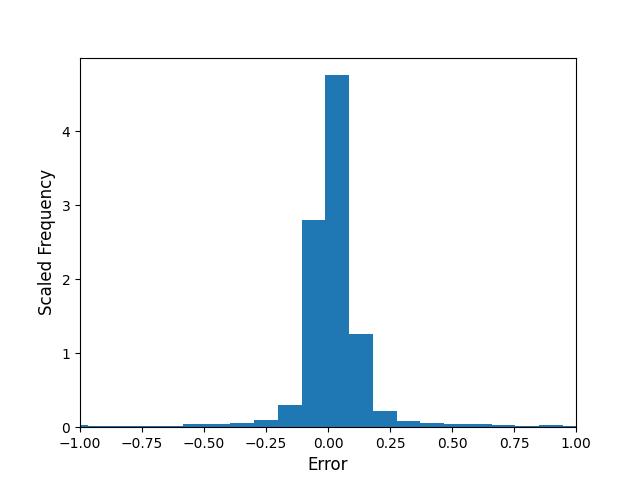}{0.5\textwidth}{}}
    \caption{
    The error distribution of the predicted MNIST digit values of 10,000 test data points from our model.\label{fig: mnist_errors}}
\end{figure}
MNIST is a large database, comprised of 70,000, $28\times 28$ images of hand-written digits \citep{deng2012mnist}. As a first step, we use this data set to test our architecture. The labeled interpolator we used for this dataset, both the encoder and the decoder, is a simpler version of what is shown in \autoref{fig: BH Encoder} and \autoref{fig: BH Decoder}. We trained the model on 50,000 images and used 10,000 images for validation. We set aside 10,000 for testing. 

We chose the latent space dimension to be equal to 10. We chose a fixed learning rate of $10^{-3}$ for all epochs. After training for 1 epoch, we sampled randomly from the test images and chose one image for each digit. We tested the model on the 10 sample images. We repeated this process for each epoch for 50 epochs. The results can be seen in \autoref{fig: MNIST epochs}. From the last epoch, it can be seen that we have indeed reconstructed the original image to a good approximation.

Furthermore, for some epochs, ambiguity in digit images produces a corresponding ambiguity in parameter estimates. For example, if the reconstructed image of the digit resembles another digit, the predicted number from the second branch of ALINet would be a number between the two digits, almost equal to the average (\autoref{fig: MNIST epochs}). This shared ambiguity between the two decoder branches provides evidence for the existence of a strong correlation between the generated images and their corresponding parameters.

Prediction error distribution of the MNIST-trained model is plotted in \autoref{fig: mnist_errors}. 
With a $1\sigma$ error less than $\pm0.25$, it is immediately evident that the ALINet predictions provide a good approximation for the actual digit value.  Taking the integer part, a procedure permitted here by the quantized nature of the underlying parameter (digit value), recovers exactly the true digit values.
\subsection{Black Hole Model: Parameters From Image}\label{subsec: ALINet im to p}
\begin{figure*}[!ht]
    \fig{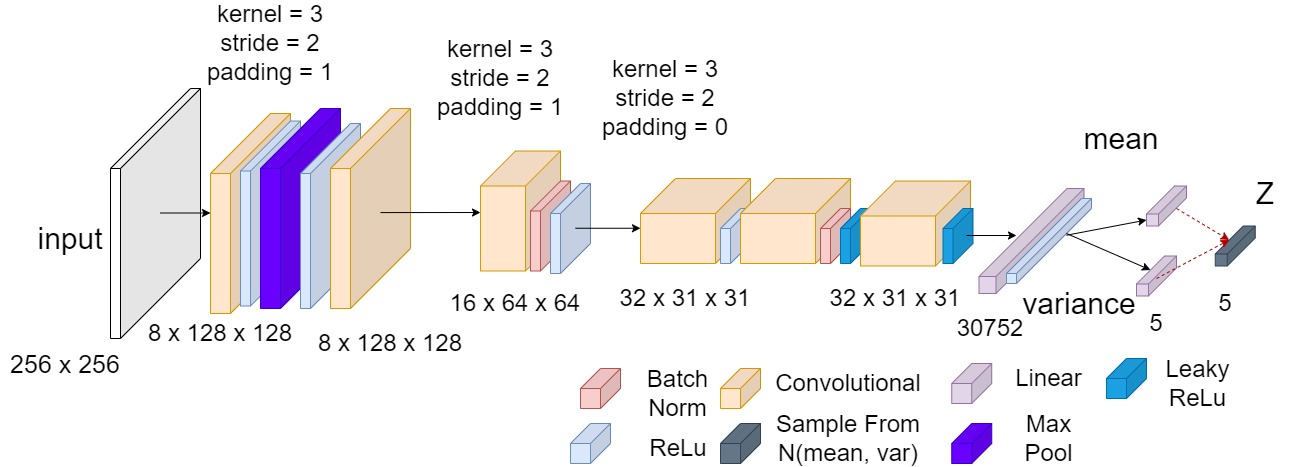}{0.9\textwidth}{Encoder}\label{fig: BH Encoder}
    \fig{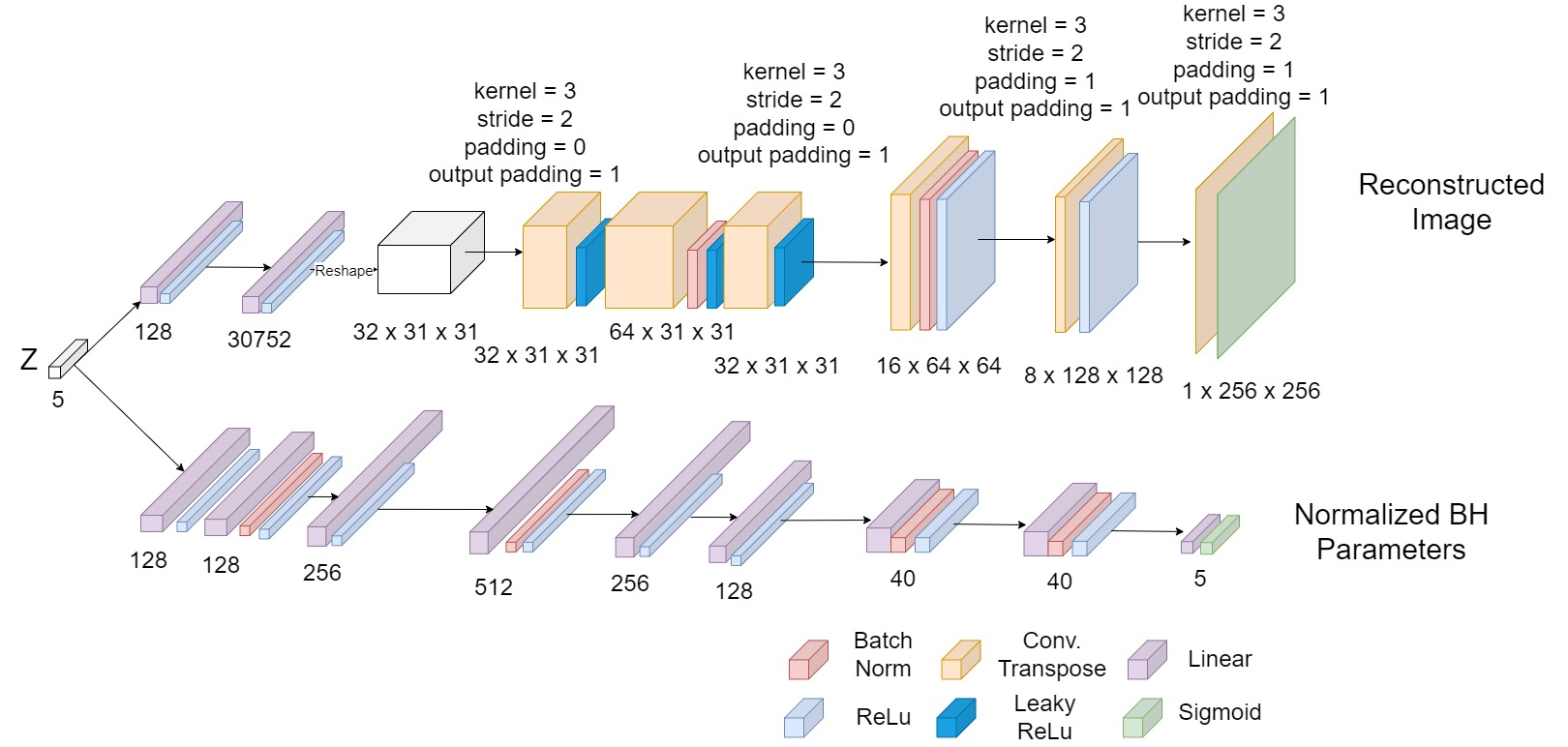}{\textwidth}{ALINet Decoder}\label{fig: BH Decoder}
    \caption{
    The ALINet architecture used for the black hole model. Note that the numerical output of the second branch is bounded between 0 and 1 due to final Sigmoid layer.
    To recover the physical parameters, we must ``unnormalize'' them, which can be done with the prior knowledge of their individual ranges, listed in \autoref{table: parameter ranges}. \label{fig: 200k architecture}}
\end{figure*}
\begin{deluxetable*}{lccccB}[p]
\tabletypesize{\scriptsize}
\tablewidth{0pt} 
\tablecaption{Physical parameter ranges for the RIAF image model. \label{table: parameter ranges}}
\tablehead{
 \colhead{physical parameter} & \colhead{description}& \colhead{minimum} & \colhead{maximum} & \colhead{number of samples from the range}
}
\startdata 
Black hole spin & $a$ &  $-0.98$ & $0.98$ & $10$\\
Cosine of inclination angle & $\mu$ &  $-0.99$ & $0.99$ & $10$\\
Height-to-radius ratio of accretion disk & $H/R$ &  $0.05$ & $2$ & $10$\\
Proper number density of electrons in accretion disk & $n_{\rm nth}$ &  $0$ & $0.05$ & $10$\\
Subkeplerian fraction& $\kappa$ &  $0.01$ & $1$ & $10$\\
\enddata
\end{deluxetable*}
\begin{figure*}[p]
\centering
\gridline{\fig{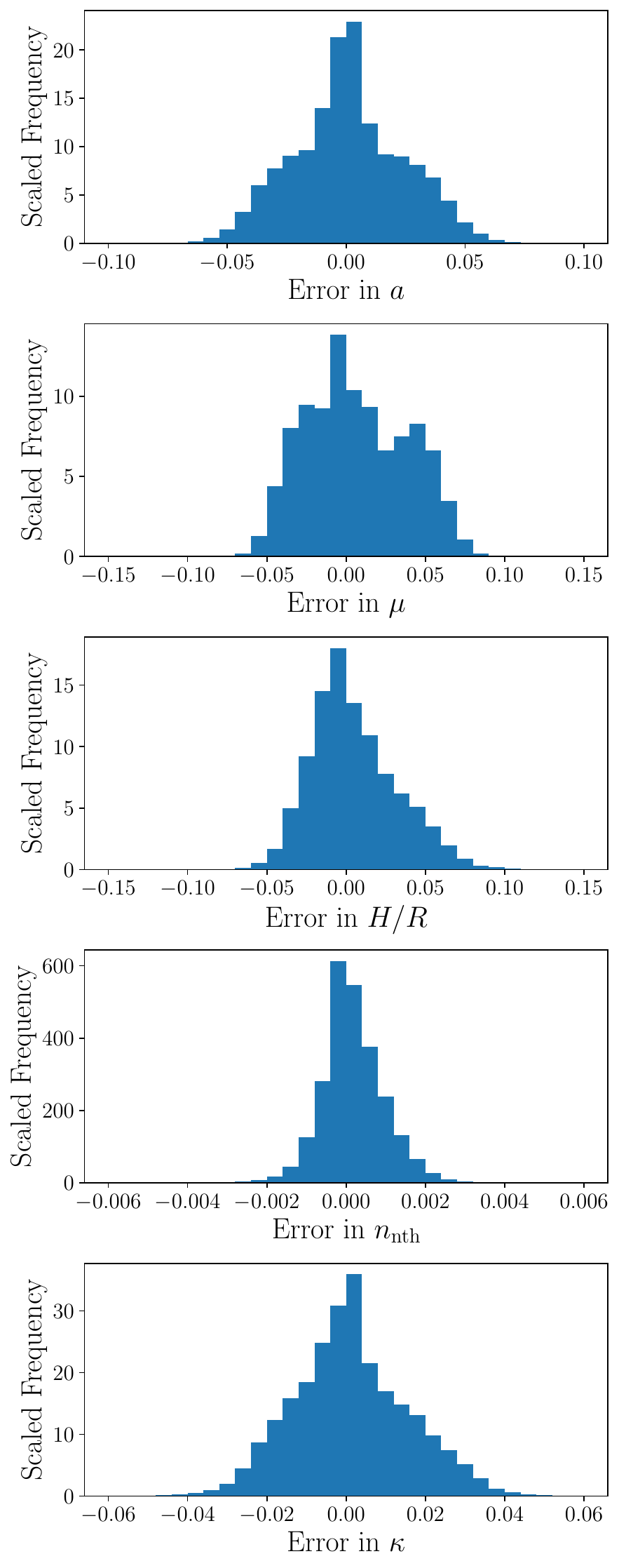}{0.45\textwidth}{}\label{subfig: ALINet_error}
\fig{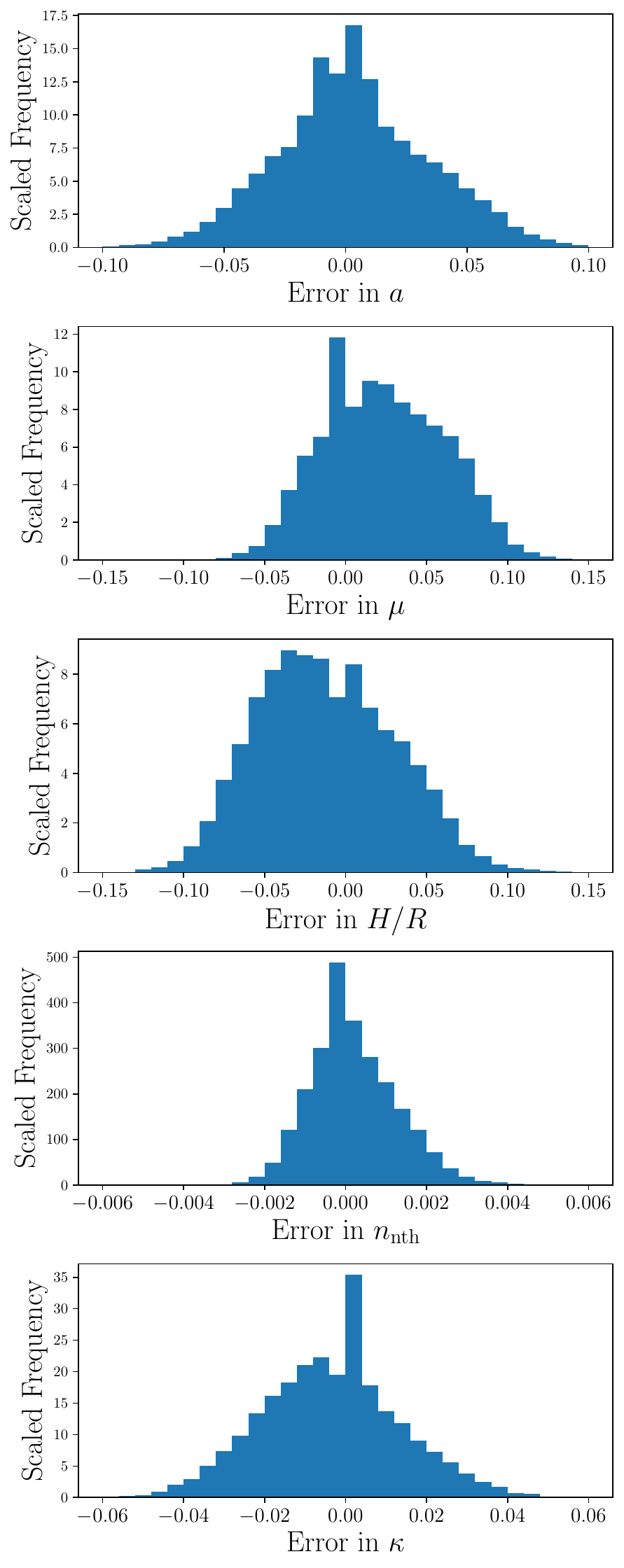}{0.45\textwidth}{}\label{subfig: InvNet_error}}
\caption{
The error distributions of the RIAF parameter values obtained from the second branch in ALINet for 20,000 test images (left) and for 20,000 test parameter values using ALINet+InvNet (right). The values against which these should be compared can be found in \autoref{table: parameter ranges}.
}
\label{fig: errors}
\end{figure*}

\begin{figure*}
\centering 
\includegraphics[width=\textwidth]{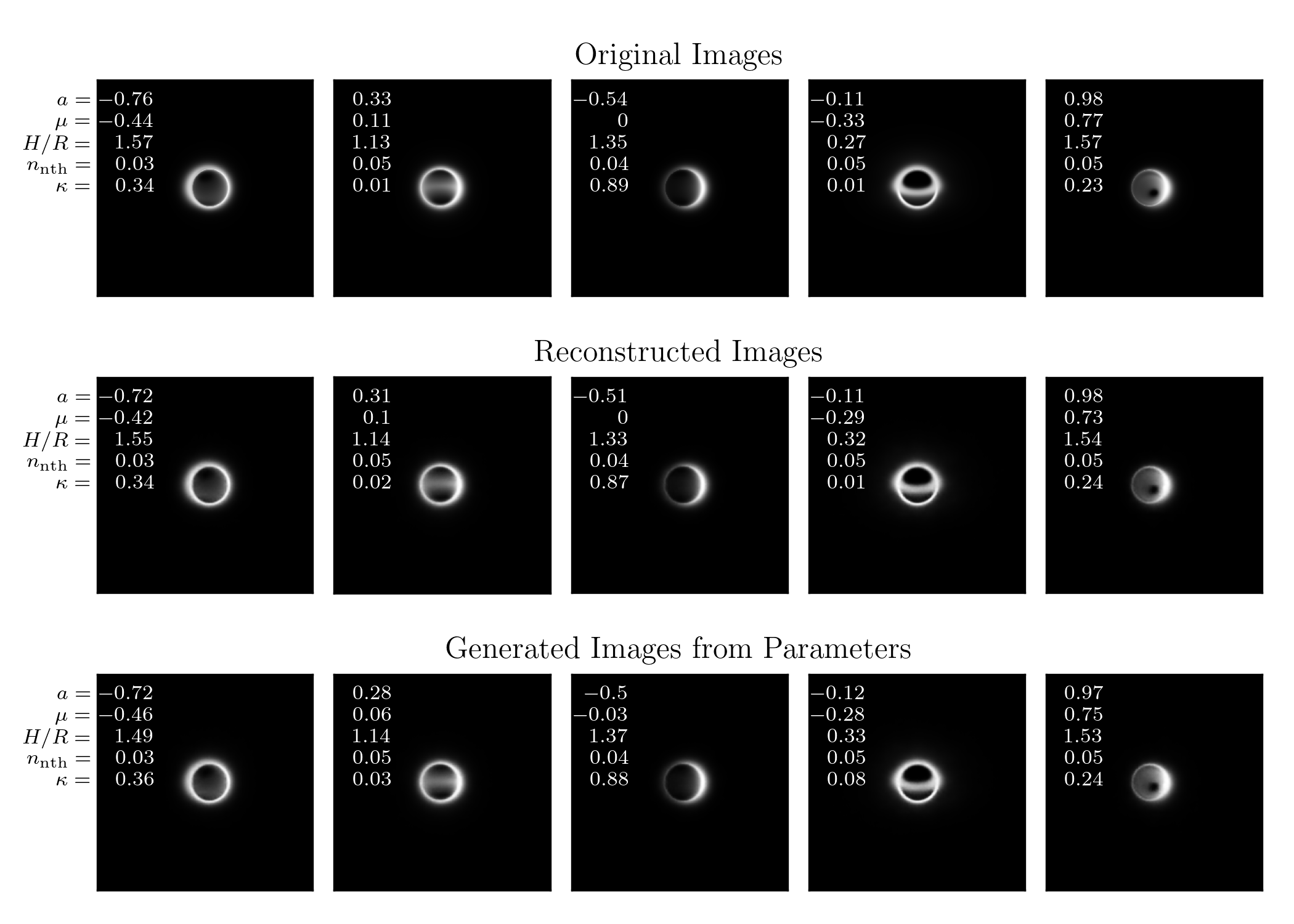}
\caption{Five sample images and their reconstructed outputs from the RIAF model. The number in each panel corresponds to the true physical parameters for that image in the first row and the reconstructed physical parameter values from the second branch of the ALINet decoder for the second row. In the third row, the images are generated by giving the truth parameter values as input to ALINet+InvNet, and the predicted parameters from the second branch of ALINet decoder are shown.}\label{fig: 200k model BH samples}
\end{figure*}
\begin{figure*}
\centering
\includegraphics[width=0.9\textwidth]{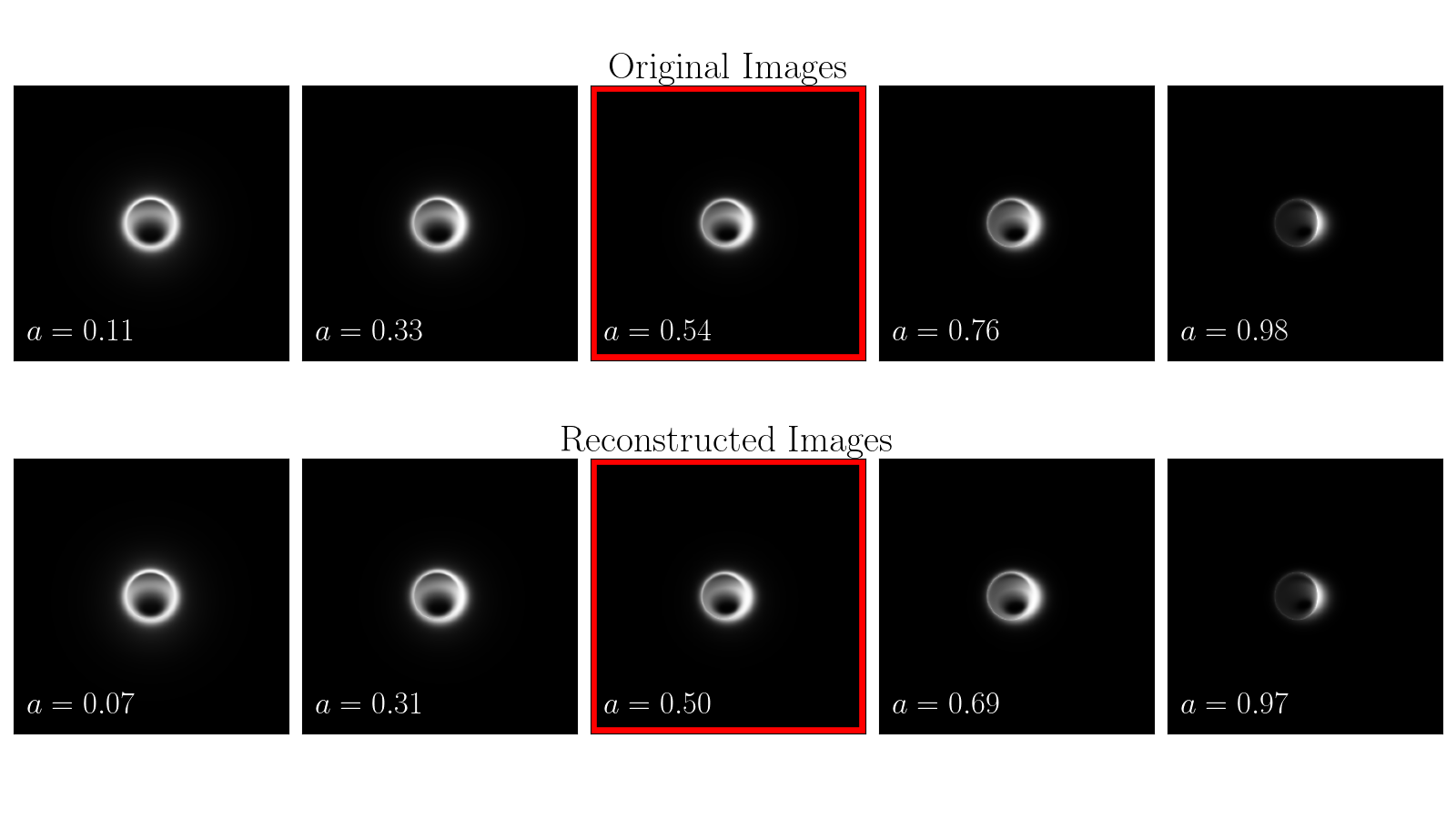}
\caption{Examples of the original and recovered images as a function of spin using InvNet+ALINet. The center image in each row, distinguished by the red boundary, is in the testing dataset and has not been seen by the network. All other parameters are fixed ($[\mu,\, H/R,\, n_{\rm nth},\, \kappa]=[-0.55,\,1.13,\, 0.028,\,0.12]$).}\label{fig: varying sping samples}
\end{figure*}
We create a library of RIAF images by sampling  a grid of physical parameter containing 10 different values from the range of each of the physical parameters. 
We choose a sufficiently dense sampling of the physical parameter space, 10 in each physical parameter, to capture all relevant morphological features in the image library.
We use 160,000 RIAF black hole images for training, 20,000 for validation, and 20,000 for testing. We choose these images in such a way that they cover the entire parameter space (look at \autoref{table: parameter ranges} for parameter ranges).
Since the RIAF images are from a physical model with five varying parameters, we choose the dimension of the latent space to be equal to five as well, thereby reducing the chances of ALINet encoding parameters that are not physical in the latent space.

We map the physical parameters to the unit interval using their minimum and maximum values. This is a useful practice when some parameters span very different ranges. It is especially useful for learning the physical parameter $n_{\rm nth}$ since its range is small compared to the other parameters.

In \autoref{eq:ELBO-ALI}, we choose a large value for $\alpha = 4\times10^4$, sufficient to make the cost function of the low-dimensional parameter space comparable to the cost function of the high-dimensional image space.
The ALINet architecture we used to train for this data set can be found in \autoref{fig: 200k architecture}. The ALINet encoder is an augmented version of a well-known convolutional architecure, AlexNet \citep{Krizhevsky2012ImageNetCW}.\footnote{Note that there are newer architectures than AlexNet available; however, these models are originally used for classifying millions of images, where the number of model parameters needed is far above what we need. AlexNet provides enough model parameters for our dataset of 200,000 images \citep{vgg, GoogLeNet, ResNet, tan2020efficientnet}.} The augmentations are done to improve numerical stability in gradient descent calculations, and to slightly increase the number parameters. The first branch of the ALINet decoder is a mirror of the encoder, where this decoder branch is doing the exact opposite of the encoder, going from a low dimensional latent space, to a high-dimensional image space. The second branch of the ALINet decoder is a chain of fully connected layers. We add as many layers as needed to extract physical parameters as accurately as possible, while keeping the process cost-effective. There is a Sigmoid at the last layer of both branches to bring the output values to range $[0, 1]$.

The model is trained for 45 epochs with a batch\footnote{To efficiently explore optimal model parameters, we segment the training data into subsets for each optimization step. A full discussion of batch optimization can be found in \citet{BatchOptOrigPaper, BatchReviewPaper}.} size of 64.
For better convergence, we used a learning rate of $10^{-3}$ for 15 epochs, then we reduced the learning rate to $10^{-4}$ and trained again for 15 epochs, and finally changed trained for 15 epochs with a learning rate of $10^{-5}$. This was motivated by relatively high fluctuations in the first and second 15 epochs, which is a sign of overshooting in gradient descent steps.
Training on Beluga (operated by Calcul Qu\'ebec, regional partner of he Digital Research Alliance of Canada) takes 12 hours to complete, using 40 CPUs and 1 GPU.
After training, the model is tested with the 20,000 testing images, and the errors are depicted in \autoref{fig: errors}. Furthermore, we have plotted the output when our model is used to reconstruct five randomly chosen images from the test data set, the results can be found in \autoref{fig: 200k model BH samples}. 

Finally, for all the test images, the difference between the predicted value of black hole image physical parameters and the true values given to the simulation can be found in \autoref{fig: errors}. These errors, signifying physical parameter fidelity of the network, should be added as systematic errors to any further analyses.  We discuss their size relative to other sources of error in EHT image analysis in the following section.

\subsection{Black Hole Model: Image From Parameters}\label{subsec: p to im exp}
Here we generate images directly from the physical parameters by combining InvNet with ALINet. We use the same dataset as \autoref{subsec: ALINet im to p}. Note that, as explained in previous sections, an ALINet has already been previously trained on this dataset. Now, an InvNet is trained for 2 epochs with learning rate of $10^{-4}$ and then for another 2 epochs with learning rate of $10^{-5}$. In \autoref{Eq: InvNet} $\alpha=10$. The error (i.e., difference between the reconstucted and origional $P$) of the physical parameters extracted when InvNet + ALINet decoder is used is shown in \autoref{fig: errors}.  
These figures and \autoref{table: parameter ranges} show that the $1\sigma$ errors for all the parameters for both ALINet and InvNet are less than or equal to $2.5\%$, which is much smaller than typical prediction errors reported, e.g., in \citet{Broderick_2020}, where prediction error of spin in RIAF model with visibility amplitude analysis is around $20\%$, and adding closure phase analysis reduce this error down to $5\%$. 
These must be added to model misspecification errors (associated with the application of smooth, station semi-analytical models to turbulent, highly variable sources) and those due to propagation effects (e.g., refractive scattering in the interstellar medium).
Thus, the current ALINet uncertainties are not expected to significantly contribute to the error budget of physical parameter estimates with current EHT data.

Note that, in the left column of \autoref{fig: errors}, the error distributions are modestly tighter and more centered around zero, which is 
a consequence of InvNet having
been trained on the output of the already-trained ALINet encoder. The errors in the InvNet predictions are therefore an accumulation of errors in ALINet and the errors in InvNet itself. Furthermore, to show that our model is achieving the task of image interpolation with high image fidelity, we vary images reconstructed from the decoder by varying one of the parameters (\autoref{fig: varying sping samples}). By varying the spin and creating the images, we can see that the reconstructed images are accurate representations of the simulation images, thereby showing high image fidelity of ALINet. Furthermore, the third row of \autoref{fig: 200k model BH samples}, we generate images using InvNet + ALINet. These images are generated by using the parameter values of the first row of this figure as input to InvNet + ALINet. The second branch of ALINet also outputs the physical parameter corresponding to each image, which match the input within a small error margin.

\section{Conclusion}\label{sec: conclusions}
In this paper we develop an interpolation tool, ALINet and ALINet+InvNet, which provide convenient and accurate machine learning tools for the construction of images of physical models appropriate for EHT data analysis from known physical parameters. We modify a very well-known machine learning architecture, a variational autoencoder, so that not only does it continuously encode images in a low dimensional space and recovers the image from this low-dimensional representation, it can also retrieve the parameters that explain the underlying behavior of each image.

We find that our ALINet model can successfully reconstruct images of hand-written set of digits, i.e., MNIST dataset, with high fidelity, while recovering their most important defining parameter, the digit represented in the image. By testing the model on a set of testing images, we showed that the $1\sigma$ prediction error of our model is about $0.1$, small enough that when the predicted value is rounded to the closest integer, the obtained value and the truth value are identical in almost all of the cases.

We also show that ALINet works well when trained on a set of black hole RIAF images. It can recover the five underlying physical parameters of all images to within $2\%$ accuracy. We show that the distributions of errors on 20,000 testing images are Gaussians centered around zero, and their deviation from zero is relatively small, thereby proving that the model is robust in extracting the black hole physical parameters. This robustness means that generating an image by giving a set of physical parameters to the ALINet+InvNet is essentially equivalent to generating an image using the ray-tracing simulations which require the computationally-expensive task of solving radiative transfer equations. Our method, therefore, increases the efficiency with which images can be generated, such that the time required to create an image from simulations on a powerful supercomputer is reduced from a few minutes to a few milliseconds.
Furthermore, since we have a mapping from the latent space to the underlying parameters of each image, the problem of interpretability of the latent space is resolved. 

ALINET+InvNet provides an efficient image model that makes physical parameter estimation via existing posterior exploration tools (e.g., \themis) from observations by EHT and future arrays.  Validation, calibration and the utilization of the RIAF image model presented here will be reported in future publications.

While the current model is similar to that used previously by \citet{Broderick_2016}, it is by no means unique.  Additional physical parameters and/or alternative underlying models that capable of producing deterministic models with smoothly varying images can be used to generate libraries from which ALINet and ALINet+InvNet models can be trained.  Generically, this process (training + parameter estimate) will be significantly faster than direct image generation via integrating the equations of radiative transfer.

Finally, this method is designed to facilitate fitting physical model parameters to EHT observational data. Since the cost of creating an image is very small, about 1~ms to create an image with one Tesla T4 GPU \citep{cuda} in Python and almost the same amount of time with only CPUs in \CC, creating a large library of images and fitting parameter distributions to the observational data is now possible.

\begin{acknowledgments}
    We thank Boris Georgiev and Sebastian Wetzel for helpful comments.
    This work was supported in part by Perimeter Institute for Theoretical Physics.  Research at Perimeter Institute is supported by the Government of Canada through the Department of Innovation, Science and Economic Development Canada and by the Province of Ontario through the Ministry of Economic Development, Job Creation and Trade.
    A.E.B. receives additional financial support from the Natural Sciences and Engineering Research Council of Canada through a Discovery Grant.
    This research was enabled in part by support provided by Calcul Qu\'ebec (www.calculquebec.ca) and the Digital Research Alliance of Canada (alliancecan.ca).
\end{acknowledgments}

\clearpage
\bibliography{main}{}
\bibliographystyle{aasjournal}

\appendix
\section{Summary of ALINet Architectural Components}
The architecture of ALINet as shown in \autoref{fig: 200k architecture} is presented as a sequence of individual (potentially nonlinear) operations that are commonly used in machine learning applications.  Here we provide a short conceptual summary of each of these.
\begin{itemize}
    \item \textbf{Linear [Layer]}: This layer applies a weight matrix $\textbf{w}$ and bias vector $b$ to an input vector such that the output is,
    \begin{equation}
        \text{output} = \textbf{w}\,.\,\text{input} + \textbf{b},
    \end{equation}
    which is equivalent to linearly transforming every single element in the input vector.
    \item \textbf{Convolutional [Layer]}: This layer slides kernels over the input with a specified stride, i.e., convolving the inputs with kernels, and thereby reducing their dimensions after the convolutional layer. The general formula for the dimension of output of this layer is,
    \begin{equation}
        d_{\text{out}} = \lfloor \frac{(d_{\text{input}} - k + 2 \times \text{padding})}{\text{stride}} \rfloor + 1,
    \end{equation}
    where $k$ is the kernel size, padding is the dimension of the padding (zero pixels) added to the edges of the input matrix, stride is the size of the step the kernel takes when convolved with the input.
    \item \textbf{Transposed Convolutional Layer (Conv. Transpose)}: This is the opposite of a convolutional layer. The output from this layer has bigger dimensions than the input, where the output dimension is determined from 
    \begin{equation}
        \begin{aligned}
            d_{\text{out}} &= (d_{\text{input}} - 1) \times \text{stride} + k - 2 \times \text{padding}\\
            &+ \text{output\_padding} + 1.
        \end{aligned}
    \end{equation}
    where output\_padding is the extra padding we manually add to the output of the Conv. Transpose layer, and stride, padding, and $k$ are all equal to the counterpart convolutional layer hyperparameters that this Conv. Transpose layer is trying to reverse.
    \item \textbf{Activation Function}: Functions applied to every element of a matrix or vector that is the output of a linear, convolutional, or transposed convolutional layer. These functions are used to add non-linearity to the functions learned by the neural network. Different functions are used as activation functions; here are the three we used in this paper:
    \begin{itemize}
        \item \textbf{ReLU}: This function is linear for positive argument and vanishes otherwise, i.e., 
    \begin{equation}
        \text{ReLU}(x) = 
        \begin{cases}
            x & \text{if}~x\ge0\\
            0 & \text{otherwise.}
        \end{cases}
    \end{equation}
    \item \textbf{Leaky ReLU (LReLU)}: This function is similar to ReLU, but for negative argument returns a small biased value to facilitate better gradient performance, 
    \begin{equation}
        \text{LReLU}(x) = 
        \begin{cases}
            x & \text{if}~x\ge0\\
            \alpha x & \text{otherwise.}
        \end{cases}
    \end{equation}
    where $\alpha$ is chosen to be a very small number.  Because it is a hyperparameter, it must be explicitly set for individual problems.
    \item \textbf{Sigmoid}:
    This function 
    is used to smoothly normalize unbounded input to the range $[0,1]$,
    \begin{equation}
        \text{Sigmoid}(x) = \frac{1}{1 + e^{-x}}.
    \end{equation}
    \end{itemize}
    \item \textbf{Batch Normalization (Batch Norm)}: This layer is added to increase stability in calculating the gradients in the neural network. Assume $x$ is the current input training batch. Then the output of the batch normalization layer is the normalized version $\hat{x}$ where we have:
    \begin{equation}
        \hat{x}_i = \frac{x_i - \mu_i}{\sqrt{\sigma_i^2 + \epsilon}},
    \end{equation}
    where $x_i$ is the i-th element in vector $x$ and epsilon is added to the denominator for numerical stability.
    \item \textbf{Max Pool [Layer]}: This layer is added to facilitate a small amount of translation invariance, i.e., translating the image by a small amount does not change significantly the output of the max pool layer. Max pooling calculates the maximum value of patches of the input, and replaces that patch with that maximum value.
\end{itemize}
\end{document}